\documentclass{pasa}%

\usepackage{graphicx}
\usepackage{pdflscape}

\title[Mass of prominences]{Mass of prominences experiencing failed eruptions}

\author[B. Filippov]{B. Filippov \thanks{E-mail:
bfilip@izmiran.ru}
\affil{Pushkov Institute of Terrestrial Magnetism,
Ionosphere and Radio Wave Propagation of the Russian Academy of
Sciences (IZMIRAN), \\ Troitsk, Moscow 108840, Russia}%

}%

\jid{PASA}
\doi{10.1017/pas.\the\year.xxx}
\jyear{\the\year}

\usepackage{aas_macros}
\usepackage{hyperref} 
\hypersetup{colorlinks,citecolor=blue,linkcolor=blue,urlcolor=blue}

\hypersetup{draft}

\begin{document}

\begin{frontmatter}
\maketitle

\begin{abstract}
A number of solar filaments/prominences demonstrate failed eruptions, when a filament at first suddenly starts to ascend and then decelerates and stops at some greater height in the corona. The mechanism of the termination of eruptions is not clear yet. One of the confining forces able to stop the eruption is the gravity force. Using a simple model of a partial current-carrying torus loop anchored to the photosphere and photospheric magnetic field measurements as the boundary condition for the potential magnetic field extrapolation into the corona, we estimated masses of 15 eruptive filaments. The values of the filament mass show rather wide distribution in the range of $4\times10^{15}$ -- $270\times10^{16}$ g. Masses of the most of filaments, laying in the middle of the range, are in accordance with estimations made earlier on the basis of spectroscopic and white-light observations.   
\end{abstract}

\end{frontmatter}

\section{INTRODUCTION }
\label{sec:intro}

Early models of prominence support were aimed at the search of an agent able to withstand the gravity force high in the corona. It was magnetic field that was considered as most suitable to resolve this problem \citep{Me51,Ki57,Ku74}. Later there were found magnetic configurations that can provide equilibrium of coronal structures without taking into account their weight, and that this equilibrium can be suddenly lost leading to eruption \citep{va78,Mo87,Pr90,Fo91,Li98,Sc13}. On the other hand, resent researches showed that mass may be able to influence the local and global properties of coronal magnetic configuration \citep{Lo03,Pe07,Se11,Gu13,Bi14,Re17,Je18,Ts19}. Therefore, estimation of prominence mass and its role in equilibrium and initiation of an eruption is significant for theory, modeling, and prediction of eruptive events on the Sun. 

There are several methods of measuring prominence mass. Mass estimation using typical prominence neutral hydrogen density and geometrical dimensions gives the total prominence mass $M$ in the range $5\times10^{12} - 10^{15}$ g \citep{La10}. Measurements of continuum absorption observed in EUV yield total mass values ranging from $10^{14}$ to $2\times10^{15}$ g \citep{He03,Gi06,Gi11}.  \citet{Ca16} estimated the mass of the intermediate filament before an eruption on 2015 March 15 as $2.4\times10^{15}$ g. Prominence mass can be calculated on the base of cloud models by comparing observations with the results of non-LTE magneto-hydrostatic models. This method gives values in the range $2\times10^{15} - 6\times10^{15}$ g \citep{Ko08,Gr14}. White light coronagraph observations are used to calculate the contribution of erupting prominences to the mass of coronal mass ejections (CMEs). \citet{At86} took into account both white light and H$\alpha$ emissions and obtained a mass of $10^{16}$ g for the eruptive prominence on 1980 August 18. Some authors reported even greater values of eruptive prominence masses. \citet{Ru82} estimated the mass of the quiescent prominence before the eruption on 1980 August 18 as $2.5\times10^{16}$ g. \citet{Go98} determined the mass of the quiescent prominence, which erupted on 1994 April 4, as $6\times10^{16}$ g. It should be noted that all evaluations of prominence masses strongly depend on several poorly known parameters such as ionization degree, filling factor, optical thickness, chemical composition, etc. These uncertainties together with difficulties in the interpretation of the prominence spectrum permit only order of magnitude estimates to be made.

Different approach to estimate the prominence mass used \citet{Lo03}. They treated solar quiescent prominences as a thin plasma sheet suspended in an axisymmetric, hydromagnetic atmosphere. They suggested that the magnetic energy needed for driving CMEs is stored prior to an eruption in a magnetic flux rope held in equilibrium by the weight of a quiescent prominence. For coronal fields of 5–10 G, hydromagnetic solutions suggest that a prominence mass of $10^{16} - 26\times10^{16}$ g is needed to hold detached magnetic fields of intensity comparable to the coronal fields. 

In this paper, we estimate mass of prominences experiencing failed eruptions with the help of a model of a partial current-carrying torus loop anchored to the photosphere. We suppose that in most considered cases the gravity force plays a major role in the confinement of the eruptions. However there are examples showing other reasons for the termination of eruptions.

\begin{figure*}
		\includegraphics[width=180mm]{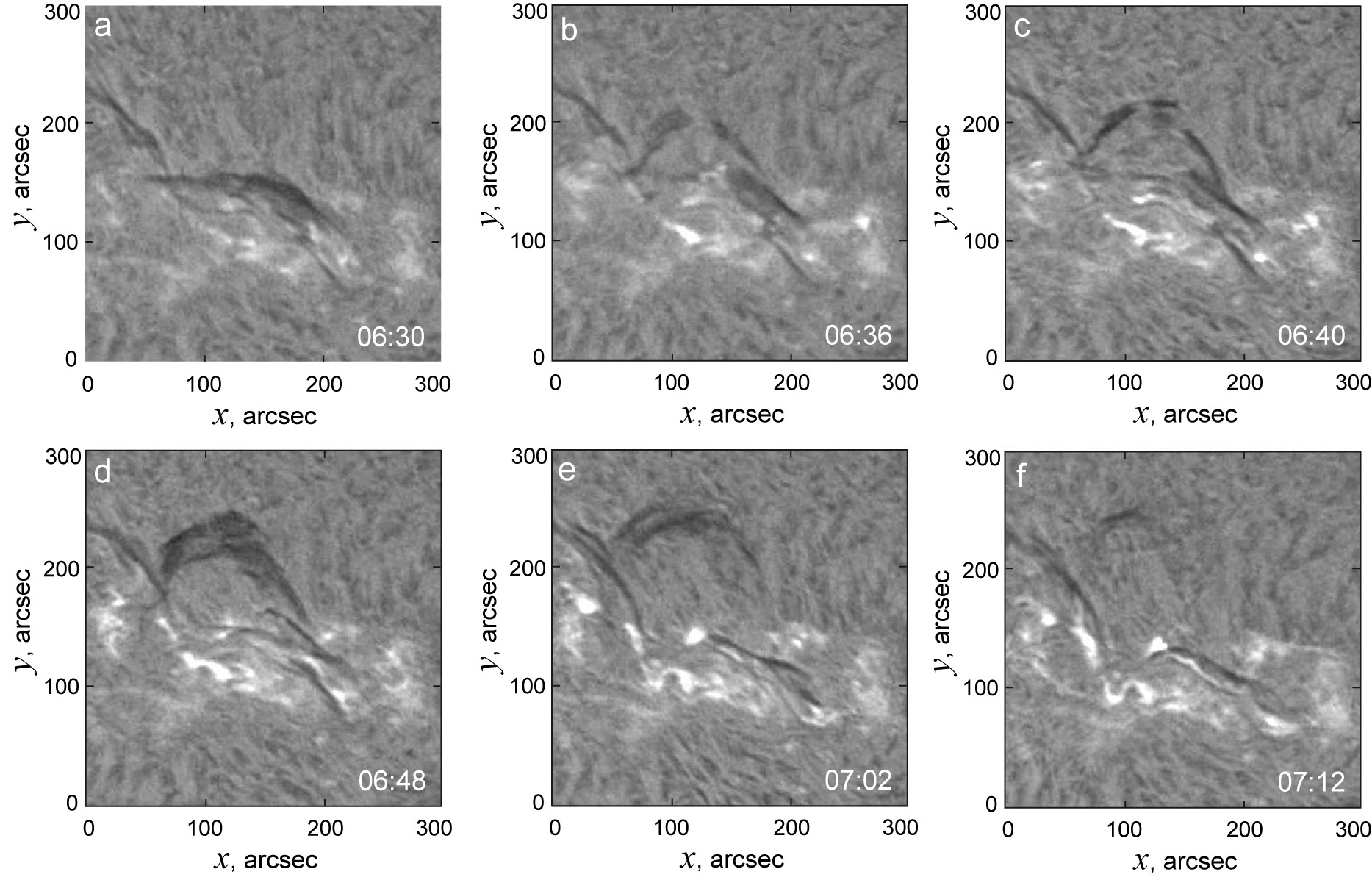}
    \caption{ H$\alpha$ filtergrams showing the failed filament eruption on 2014 March 20. (Courtesy of the Udaipur Solar Observatory.) }
    \label{Fig1}
\end{figure*}

\section{Observational data}

We selected a number of confined filament eruptions from the AIA Filament Eruption Catalog (http://aia.cfa.harvard.edu/filament/) \citep{Mc15}  observed by the Atmospheric Imaging Assembly [AIA: \citep{Le12}] onboard the {\it Solar Dynamics Observatory} [{\it SDO}: \citep{Pe12}]. Since eruptive events are more frequent at solar maximum, most events belong to the maximum phase.

It is evident that the necessary condition is the absence of the associated CME. The need to measure the final height of the filament above the photosphere prescribed the selection of the events that were not too close to the centre of the solar disk, however in some cases the calculation of the neutral surface can be helpful for the estimation of height \citep{Fi16}. Events located close to the east limb were not suitable, because the initial filament position relative photospheric magnetic fields would be difficult to determine. We selected 15 events observed during the period from May 2013 to July 2014 and studied in a paper by \citet{Fi20b} (hereafter referred as Paper I). For comparison, we also consider several successful eruptions. 

H$\alpha$ filtergrams from the Big Bear Solar Observatory, the Kanzelhohe Solar Observatory, the Udaipur Solar Observatory, and the National Solar Observatory (NSO)/Global Oscillation Network Group (GONG) were used in addition to the AIA data. Observation in EUV with the Sun Earth Connection Coronal and Heliospheric Investigation (SECCHI) Extreme Ultraviolet Imager [EUVI: \citep{Wu04,Ho08}] ) onboard the {\it Solar Terrestrial Relations Observatory} ({\it STEREO}) allowed the measurements of the final eruptive prominence height and in some cases its initial height before the eruption if a filament was close to the limb for one of the {\it STEREO} spacecraft. For the potential magnetic field calculations, magnetograms taken by the Heliospheric and Magnetic Imager [HMI: \citep{Sc12}] onboard the {\it SDO} were used as the boundary conditions.

\begin{figure*}
		\includegraphics[width=180mm]{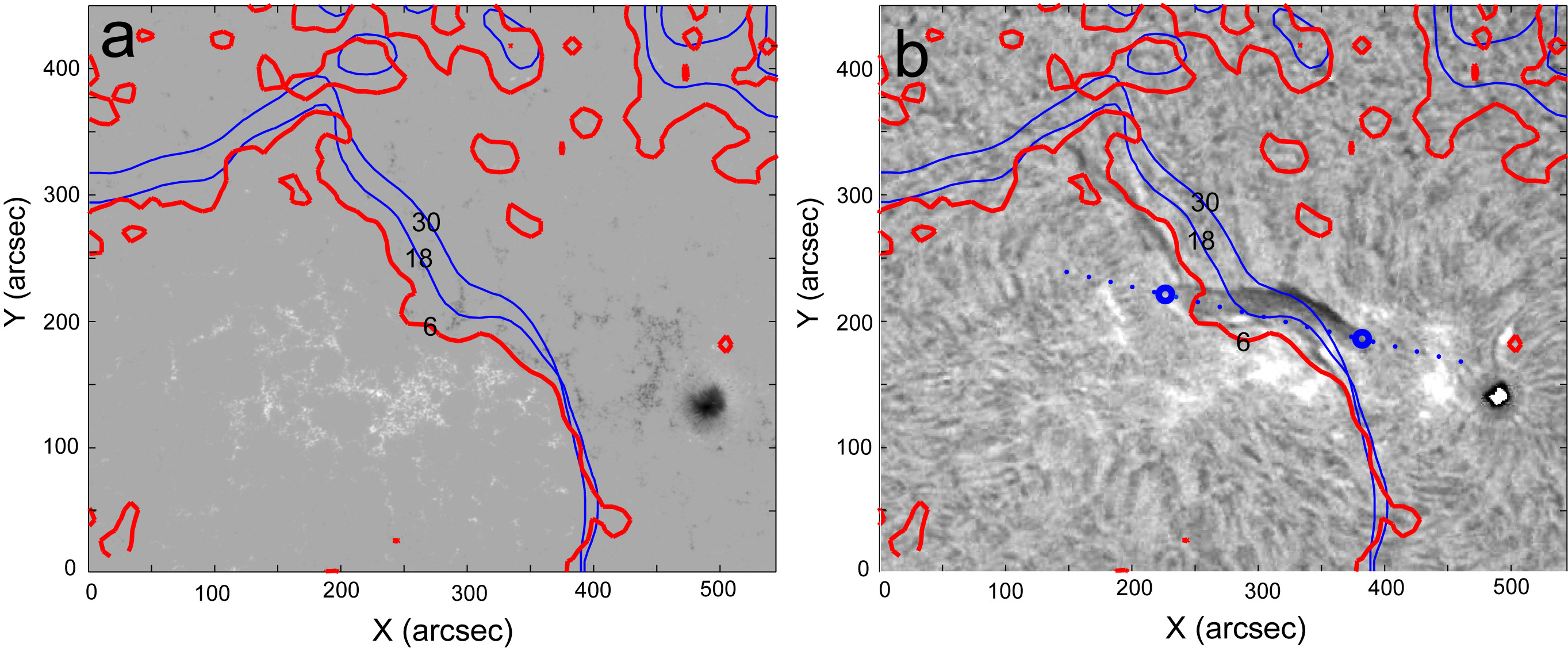}
    \caption{Fragment of the {\it SDO}/HMI magnetogram around the filament on 2014 March 20 at 06:20 UT with superposed PILs at the height of 6 Mm (thick red lines) and at 16 and 30 Mm (blue lines) (a); the same PILs superposed on the H$\alpha$ filtergram (b). Blue dots show the points above which the potential magnetic field was calculated. Blue circles indicate the endpoints of the filament and corresponding flux rope. (Courtesy of the NASA/{\it SDO} HMI science team and the Udaipur Solar Observatory.)}
    \label{Fig2}
\end{figure*}

\section{Prominence mass estimation method}

In Paper I, a simple model for the flux-rope equilibrium was analysed in which the flux rope was considered as a section of a torus with its ends embedded in the photosphere, which keeps its partly-circular shape during all evolution and remains in the vertical plane that pass through the ends. Similar models were proposed by \citet{Ch89}, \citet{Ca94},\citet{Is07},\citet{Ol10}, and some other authors. The equilibrium of the flux rope is determined by the balance of forces acting on the flux rope

\begin{equation}
F = F_R + F_I + F_B + F_g = 0.
\end{equation} 
The first term is the radial self-force per unit length acting on the circular flux rope \citep{Sh66} 
\begin{equation}
F_R = \frac{I^2}{c^2R} \left(\ln \frac{8R}{r} - \frac{3}{2}  + \frac{l_i}{2} \right) .
\end{equation}   
where $I$ is the toroidal electric current, $R$ is the major radius of the torus, $r$ is the minor radius of the torus, and $l_i$ is the internal inductance per unit length. The equilibrium along the minor radius of the torus $r$ is taken into account in Equation (2), which suggests the force-free internal magnetic field structure. We chose thereafter $l_i = 1$, which corresponds to the linear force-free internal magnetic structure \citep{Lu51,Li98,Is07}. The major radius of the torus $R$ depends of the height of the flux-rope apex $h$ and the half-footpoint-separation $a$:

\begin{equation}
R =  \frac{a^2 + h^2}{2h} .
\end{equation}  

The self-force (2) vanishes at low heights, since the axis curvature becomes large, and a nearly straight electric current does not influence itself. On the other hand, a low-height flux rope is under the strong influence of the diamagnetic photosphere creating the upward force usually modelled by acting of the mirror current\citep{Ku74} 
\begin{equation}
F_I =  \frac{I^2}{c^2h}.
\end{equation}  
Both $F_R$ and $F_I$ are directed upwards, and the flux rope is held by the external constraining poloidal magnetic field $B_e$ 
\begin{equation}
F_B =  \frac{IB_e}{c} ,
\end{equation}   
and the gravity force 
\begin{equation}
F_g =  m g \frac{R_\odot^2}{(h + R_\odot)^2} ,
\end{equation}   
where $m$ is the mass of the filament per unit length and $g$ is the free-fall acceleration at the level of the photosphere (at the distance of $R_\odot$ from the centre of the Sun).

The internal linear force-free field structure and the conservation of the toroidal flux within the flux rope lead to the dependence of the flux-rope radius $r$ on the current $I$ \citep{Li98}:
\begin{equation}
r =  r_0\frac{I_0}{I} ,
\end{equation}   
while changes of the electric current $I$ during flux-rope evolution in the corona are determined by the conservation of the poloidal magnetic flux between the photosphere and the flux rope:  
\begin{equation}
\Phi_p =  \Phi_I + \Phi_s = \frac{I}{c} L_e + \int_S B_e ds = \mbox{const},
\end{equation}  
where $S$ is the area between the photosphere and the flux rope and $L$ is the self-inductance of the torus fraction above the photosphere. The self -inductance of a thin circular flux rope is defined as \citep{La84}
\begin{multline}
L_e =   \int\limits_{\varphi_0}^\varphi \int\limits_{\varphi_0}^\varphi  \frac{R \cos \varphi}{2 \sin \varphi}  d\varphi^\prime d\varphi = \\ = 2 R \varphi \left(\ln\tan\frac{\varphi}{4}  + 2 \cos \frac{\varphi}{2} + \ln{ \frac{8R}{r}} - 2 \right) ,
\end{multline} 
where 
\begin{equation}
\varphi = \left\{ \begin{array}{ll}
 2 \arcsin \frac{a}{R} \:, &  h \leq a \;,
\\ & \\ 
 2 \pi - 2 \arcsin \frac{a}{R}  \:, & h > a \;,
\end{array} \right.
\end{equation}  
and
\begin{equation}
\varphi_0 \approx \frac{r}{2 R} .
\end{equation}  

The value of the flux $\Phi_s$ can be calculated numerically:
\begin{equation}
\Phi_s =   \int\limits_{-l}^l \int\limits_{z_1}^{z_2} B_e dxdz,
\end{equation}  
where $x$ is the horizontal coordinate along the line connecting the footpoints with the origin halfway between them, $z$ is the vertical coordinate with the origin at the photosphere,
\begin{equation}
l = \left\{ \begin{array}{ll}
a \:, &  h \leq a \;,
\\ & \\ 
 R \:, & h > a \;,
\end{array} \right.
\end{equation}  

\begin{equation}
z_1 = \left\{ \begin{array}{ll}
0 \:, &  |x| \leq a \;,
\\ & \\ 
 \frac{h^2-a^2-\left( (a^2+h^2)^2-4h^2x^2 \right)^{1/2}}{2h}  \:, & |x| > a \;,
\end{array} \right.
\end{equation}  

\begin{equation}
z_2=\frac{h^2-a^2+\left( (a^2+h^2)^2-4h^2x^2 \right)^{1/2}}{2h}  .
\end{equation}  

It was found in Paper I that for typical dependence of the external magnetic field $B_e$ on height $h$, the constraining force $F_B$ cannot balance the upward forces $F_R + F_I$ at a greater height after a catastrophic loss of equilibrium. The gravity force $F_g$ is significant for the termination of ascending motion in the model. However, the external field was modelled in Paper I by the field of two 2D horizontal dipoles located at different depths below the photosphere. Now we calculate the mass of a filament needed to stop eruptions in real events using real distribution of the external field in the corona in potential approximation based on the photospheric magnetic field measurements. We assume that a flux rope is initially in stable equilibrium and then experiences a catastrophic loss of equilibrium at the height $h_c$. After a failed eruption it stops at the height $h_m$. Therefore three equations:

\begin{equation}
\begin{aligned}F(h_c)=0,\\
\left. \frac{\mathrm d F}{\mathrm d h} \right|_{h_c} =0, \\
F(h_m)=0
\end{aligned}
\end{equation}  
should be solved numerically to find three unknown values $h_c, I_c$, and $m$. In so doing, changes of the electric current with height according to Equation (8) should be taken into account.

\section{Example of the failed filament eruption on 2014 March 20}

Failed filament eruption happened on 2014 March 20 not far from the centre of the solar disk (Figure 1). Coordinates of the middle of the filament were N23, W21, therefore the line-of-sight component of the photospheric magnetic field rather well represents the radial magnetic field needed for potential field calculation. The eruption terminated in about 50 minutes after beginning. Suggesting the filament rose in nearly radial direction, the final height was about 200 Mm.  

As usual, we cut a rectangular area around the filament from a full-disc magnetogram and transform it into the array with pixels of equal area. We use the modified arrays as the boundary conditions for solving the Neumann external boundary-value problem [see \citep{Fi13} and references therein]. Figure 2(a) presents a fragment of the {\it SDO}/HMI magnetogram around the filament on 2014 March 20 at 06:20 UT with superposed polarity inversion lines (PILs) at the height of 6 Mm (thick red lines) and at 16 and 30 Mm (blue lines) taking into account the projection shift. In Figure 2(b), the same PILs are superposed on the H$\alpha$ filtergram taken at the Udaipur Solar Observatory showing the filament just before the eruption. The height at which the PIL is the nearest one to the filament spine can be considered as the estimate of the height of the filament top \citep{Fi16}. Calculation of PILs at greater heights shows that in the top view the PILs become arranged one above the other, or the neutral surface is nearly radial. This fact permits to suppose that the eruptive filament also moves in the radial direction and for the estimation of its final height from the horizontal displacement, only the spherical projection effect should be taken into account. 

The ends of the filament marked by the blue circles determine the flux-rope footpoint separation $2a$ = 120 Mm. We draw a line through this two points and calculate the horizontal potential magnetic field $B_e$ perpendicular to the vertical plane that passes through this line at different heights. Actually, we calculate the vertical distribution of $B_e$ above 19 points equally spread along the line within the interval of $4a$, thus covering the distance of $a$ away from either footpoint. Then values of $B_e$ in any point of the vertical plane passing through the line are found by interpolation or extrapolation. We assume that the erupting flux rope remains in the vertical plane all the time. 

Figure 3 shows the vertical profiles of the parameters of the potential magnetic field averaged over the interval of $a$ around the central point. The magnetic decay index $n$ defined as usual \citep{Ba78,Fi00,Fi01,Kl06}
\begin{equation}
n = - \frac{\partial \ln B_e}{\partial \ln h}
\end{equation}
shows monotonic behaviour [Figure 3(b)], as it is typical for dipole-like field. Indeed, there is no significant patches of different polarities apart from two major areas of opposite polarity [Figure 2(a)]. While the field rotates with height, it does not change direction to opposite even at great heights [Figure 3(c)]. We use here for the decay-index calculation only the component of the horizontal potential magnetic field perpendicular to the plane of the flux rope $B_e$ in contrast to the whole horizontal field as in some other studies (e.g. in Paper I).

This values of $B_e$ are used in solving of the set of equations (16). The radius of the flux-rope cross-section is chosen as $r_0$ = 6 Mm. The solution gives $h_c$ = 18 Mm, $I_c$ = $4\times10^{11}$ A, $m$ = $10^7$ g cm$^{-1}$. The values of the decay index at the heights $h_c$ and $h_m$ are $n_c$ = 0.43 and $n_m$ = 2.5. Changes of the electric current according to Equation (8) are presented in Figure 3(d). The current value decreases below the point of catastrophic loss of equilibrium $h_c$, increases a little above it, and declines significantly at greater heights. Figure 4 demonstrates the height dependence of the total force acting on the flux rope and the contributions to it according Equations (2), (4), (5), and (6). The curve of the total force $F$ touches the abscissa axis at $h_c$ demonstrating an unstable equilibrium and crosses the axis at $h_m$ from up to down evidencing a stable equilibrium. It should be noted that in the absence of the gravity force, the curve would approach to abscissa not crossing it, which implies a successful eruption. Of course, all obtained values depend on the assumptions taken in the model and can be considered only as estimates. On account of many uncertainties in data, all obtained values have errors no less than 50\%.

To illustrate the process of failed eruption, we solve the equation of motion
\begin{equation}
M\frac{\mathrm d^2 z}{\mathrm d t^2}  = F  + F_d ,
\end{equation}    
where $F_d$ is the drag force, which is evidently needed to prevent long oscillations of the flux rope about the upper equilibrium point or even a successful eruption, if the gained kinetic energy is sufficient to overcome the confining forces. The drag force in hydrodynamics is assumed to be proportional to the first degree of velocity for small values of the Reynolds number, $F_d = - k_1 v$, and proportional to the second degree of velocity for great values of the Reynolds number, $F_d = - k_2 v^2$ \citep{La87}. We chose the values of the coefficients by the trial-and-error method in order to have only small oscillations of the flux rope after reaching the upper equilibrium position. Figure 5 shows the height-time and velocity-time plots for $k_1 = 1.8\times10^4$ g cm$^{-1}$ s$^{-1}$, $k_2 = 1.1\times10^{-2}$ g cm$^{-2}$, and $F_d = 0$. The acceleration time needed to reach the maximum velocity is approximately the same in all cases and comprises about 15 min. However the ascending time needed to reach the maximum height is different. For linear dependence of the drag force on velocity, it is about 30 min, for squared dependence it is about 1 hour, and in the absence of the drag force the flux rope does not stop at all, although its speed is decreasing. The duration of the fast motion in the {\it SDO}/AIA observations is also about 1 hour.

\begin{figure*}
		\includegraphics[width=175mm]{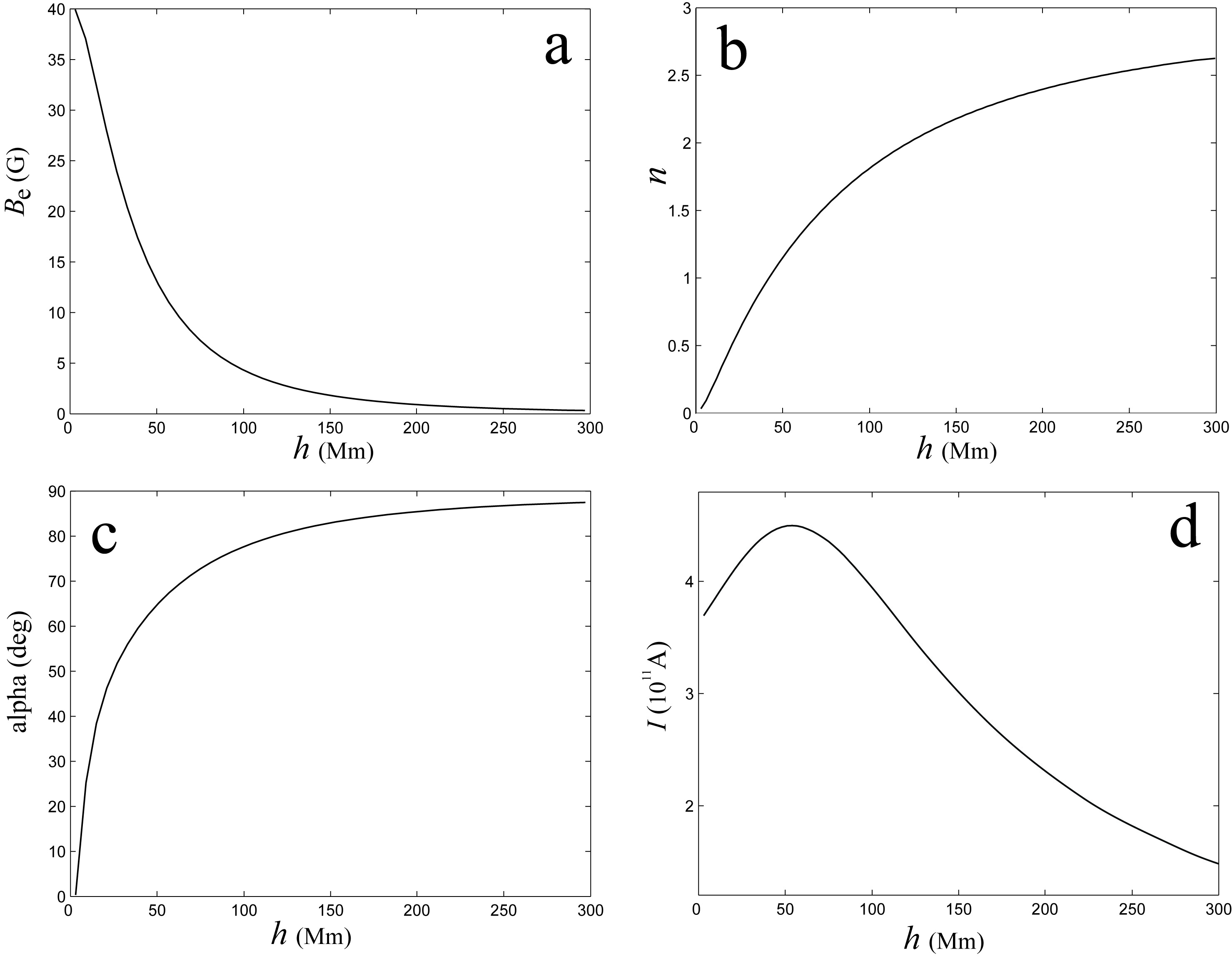}
    \caption{Dependence on height of the horizontal potential magnetic field $B_e$ (a), the decay index $n$ (b), the rotation angle $\alpha$ of $B_e$ above the middle of the filament (c), and the value of electric current $I$ in the flux rope (d). }
    \label{Fig3}
\end{figure*}

\begin{landscape}
\begin{table}
\caption{Failed filament eruptions in the regions with the monotonic decay-index height-dependence}
\label{T1}
\begin{tabular}{@{}lcccccccccccccc}
\hline
Data & Coordinates & Beginning & End & $2a$ & $h_0$ 
& $h_m$ & $B_e$ & $\alpha_m $ & $h_c$ & $n_c$ & $n_m$ &
$I_c$ & m & M \\
& (deg) &  of the fast  & of the fast &
(Mm) & (Mm) & (Mm) & (G) & (deg) & (Mm) & & & ($10^{11}$ A) &
($10^6$ & ($10^{16}$ g) \\
& &  motion UT &  motion UT & & & & & & & & & & g cm$^{-1}$) &
\\
\hline
2013/05/21 & S15 W62 & 10:00 & 11:00
& 150 & 15 & 150 & 120 & 8 & 17 & 0.5 & 2.1 &
12 & 180 & 270 \\
2013/06/11 & S05 W35 & 20:20 & 22:00
& 180 & 50 & 120 & 4 & 20 & 34 & 0.4 & 1.3 &
1.4 & 3.2 & 6 \\
2013/07/03 & N18 W40 & 05:55 & 06:10
& 55 & 10 & 70 & 4 & 40 & 12 & 0.2 & 1.6 &
0.3 & 0.18 & 1 \\
2013/09/20 & S34 W12 & 04:05 & 04:50
& 170 & 12 & 170 & 110 & 45 & 24 & 0.6 & 1.8 &
2 & 5.4 & 9 \\
2014/02/17 & S03 W02 & 02:45 & 03:00
& 100 & 18 & 180 & 90 & 80 & 12 & 0.5 & 6.1 &
6 & 17 & 17 \\
2014/03/20 & N23 W21 & 06:27 & 07:20
& 120 & 30 & 200 & 40 & 80 & 18 & 0.4 & 2.5 &
4 & 10 & 12 \\
2014/07/05 & S12 W07 & 22:30 & 22:45
& 75 & 25 & 75 & 45 & 50 & 9 & 0.4 & 2.3 &
1.8 & 7 & 5 \\

\hline
\end{tabular}
\end{table}

\begin{table}
\caption{Failed filament eruptions in the regions with the non-monotonic decay-index height-dependence}
\label{T2}
\begin{tabular}{@{}lcccccccccccccc}
\hline
Data & Coordinates & Beginning & End & $2a$ & $h_0$ 
& $h_m$ & $B_e$ & $\alpha_m $ & $h_c$ & $n_c$ & $n_m$ &
$I_c$ & m & M \\
& (deg) &  of the fast  & of the fast &
(Mm) & (Mm) & (Mm) & (G) & (deg) & (Mm) & & & ($10^{11}$ A) &
($10^6$  & ($10^{16}$ g) \\
& &  motion UT &  motion UT & & & & & & & & & & g cm$^{-1}$)  &
\\
\hline
2013/05/14 & S12 E70 & 06:03 & 07:32
& 160 & 20 & 150 & 5 & 160 & 16 & 0.6 & - 0.8 &
0.5 & 0.3 & 0.5 \\
2013/06/12 & N33 W10 & 15:13 & 15:55
& 150 & 20 & 180 & 8 & 140 & 15 & 0.6 & 1 &
0.6 & 0.4 & 0.6 \\
2013/08/04 & N20 E61 & 06:45 & 07:00
& 120 & 10 & 130 & 3 & 170 & 40 & 0.04 & 0.8 &
0.6 & 0.3 & 0.4 \\
2013/12/08 & S03 W90 & 04:10 & 05:20
& 130 & 30 & 120 & 3 & 70 & 22 & 0.6 & 19 &
0.7 & 0.7 & 0.9 \\
2014/03/27 & S24 W60 & 12:40 & 13:30
& 110 & 15 & 140 & 11 & 210 & 12 & 0.8 & 0.3 &
1 & 1.4 & 1.5 \\
2014/03/28 & S10 W66 & 01:40 & 02:10
& 100 & 15 & 120 & 110 & 150 & 8 & 0.7 & - 1.2 &
5.8 & 44 & 44 \\
2014/05/03 & N13 E63 & 07:35 & 08:10
& 80  & 15 & 130 & 30 & 90 & 12 & 0.47 & 4.7 & 
2 & 2.7 & 2 \\
2014/07/02 & N05 W52 & 17:20 & 18:20
& 190  & 25 & 160 & 20 & 50 & 17 & 0.9 & 1.8 & 
2 & 6.4 & 12 \\

\hline
\end{tabular}
\end{table}

\end{landscape}

\begin{table*}
\caption{Successive filament eruptions in the regions with the monotonic decay-index height-dependence}
\label{T3}
\begin{tabular}{@{}lcccccccccc}
\hline
Data & Coordinates & Beginning & $2a$ & $h_0$ 
 & $B_e$ &  $h_c$ & $n_c$  &
$I_c$ & m & M \\
& (deg) &  of the fast   &
(Mm)  & (Mm) & (G)  & & & ($10^{11}$ A) &
($10^6$  & ($10^{16}$ g) \\
& &  motion UT  & & & & & & & g cm$^{-1}$) & 
\\
\hline
2013/06/22 & N25 E26 & 15:00 
& 380 & 100 & 10 & 42 &  0.6 & 
2.6 & < 2.7 & < 10 \\
2013/09/23 & S31 W60 & 08:00 
& 300 & 30 & 11 & 30 &  0.6 & 
1.4 & < 0.6 & < 1.8 \\
2016/01/26 & S22 W29 & 16:55 
& 170 & 60 & 45 & 17 &  0.44 & 
3.5 & < 2.1 & < 3.6 \\

\hline
\end{tabular}
\end{table*}

\begin{table*}
\caption{Successive filament eruptions in the regions with the non-monotonic decay-index height-dependence}
\label{T4}
\begin{tabular}{@{}lcccccccccc}
\hline
Data & Coordinates & Beginning & $2a$ & $h_0$ 
 & $B_e$ &  $h_c$ & $n_c$  &
$I_c$ & m & M \\
& (deg) &  of the fast   &
(Mm)  & (Mm) & (G)  & & & ($10^{11}$ A) &
($10^6$ & ($10^{16}$ g) \\
& &  motion UT  & & & & & & & g cm$^{-1}$) & 
\\
\hline
2012/06/23 & N10 W90 & 06:30 
& 300 & 60 & 50 & 21 &  0.6 & 
4.3 & < 3.2 & < 10 \\
2013/09/29 & N20 W24 & 21:25 
& 450 & 80 & 30 & 33 &  0.67 & 
3 & < 2 & < 10 \\
2014/03/29 & N52 E53 & 01:53 
& 70 & 20 & 65 & 7 &  0.4 & 
0.5 & < 0.3 & < 0.2 \\

\hline
\end{tabular}
\end{table*}

\begin{figure}
		\includegraphics[width=85mm]{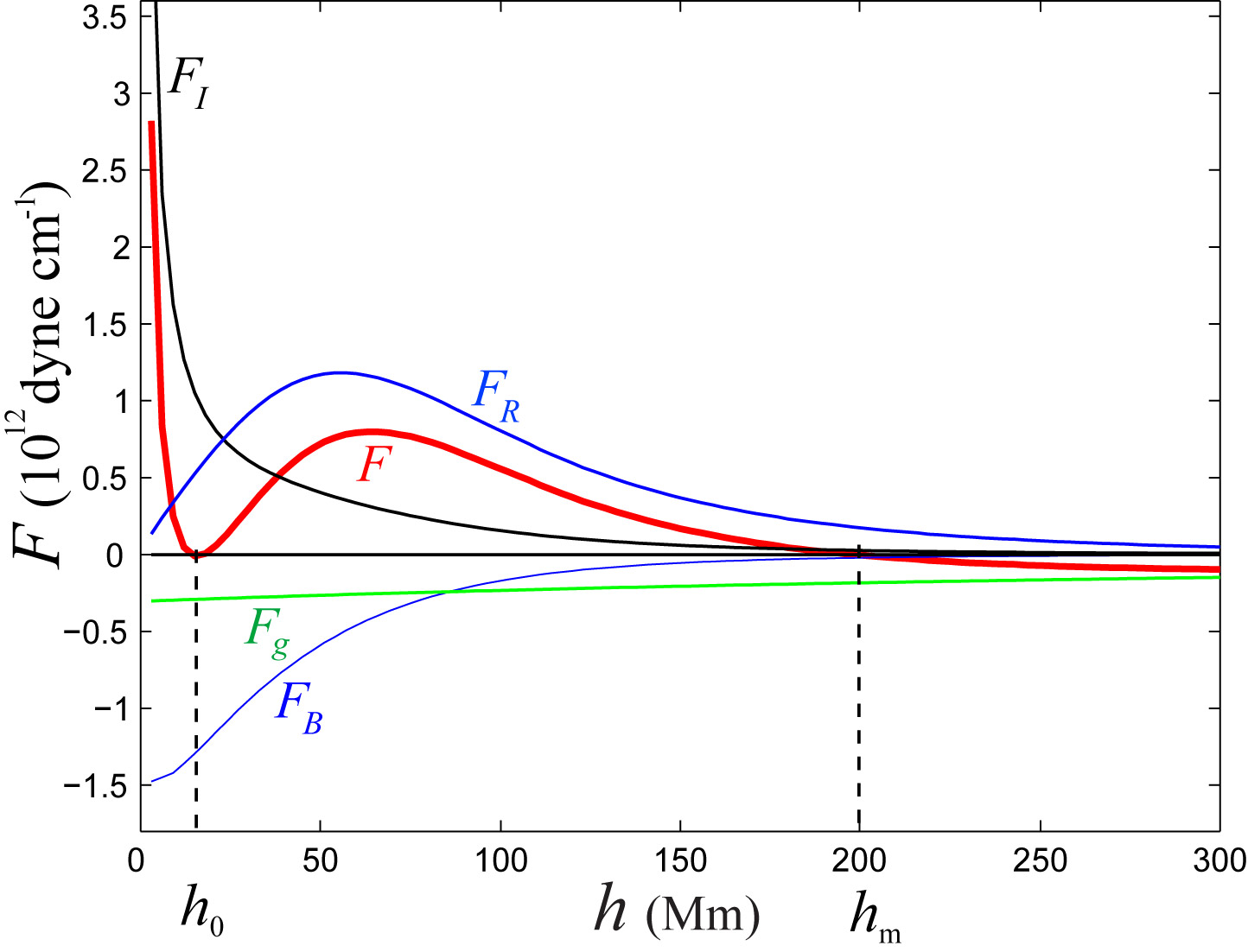}
    \caption{Vertical distributions of the forces acting on the flux rope. }
    \label{Fig4}
\end{figure}

\begin{figure*}
		\includegraphics[width=170mm]{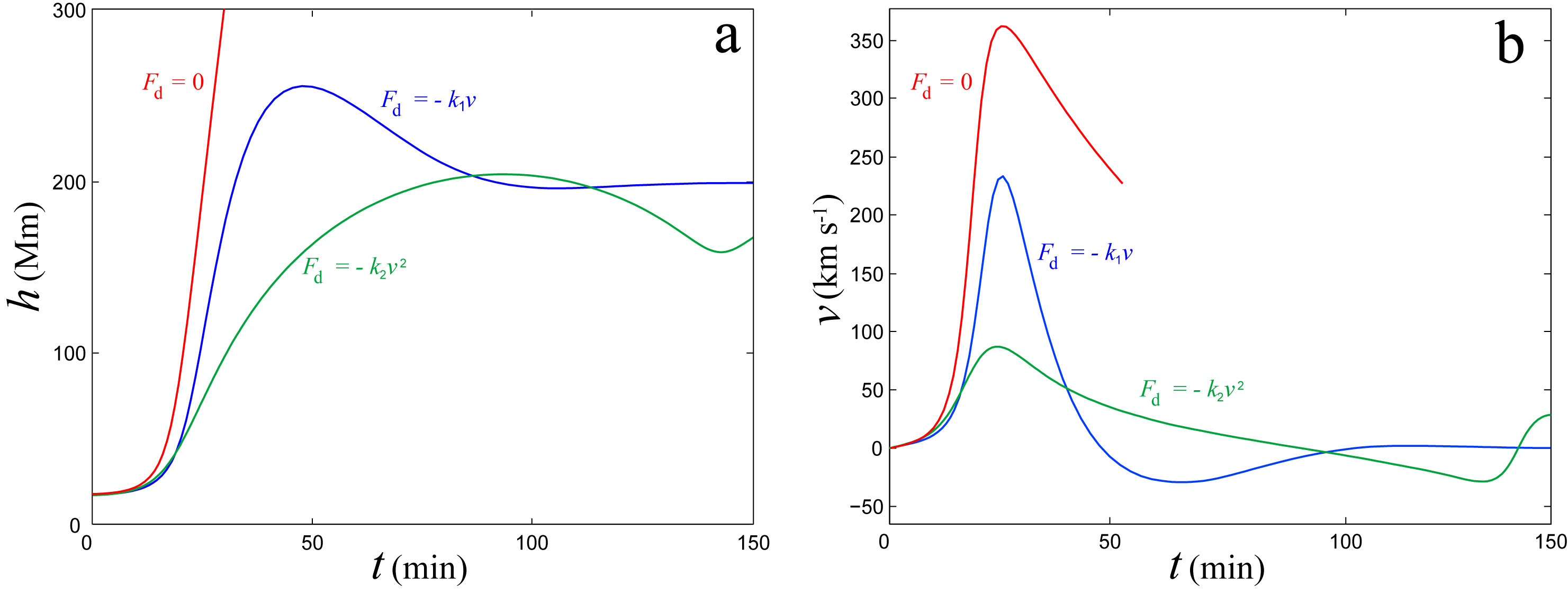}
    \caption{Height and velocity of the erupting flux rope experiencing the action different kinds of the drag force.  }
    \label{Fig5}
\end{figure*}

\begin{figure*}
		\includegraphics[width=180mm]{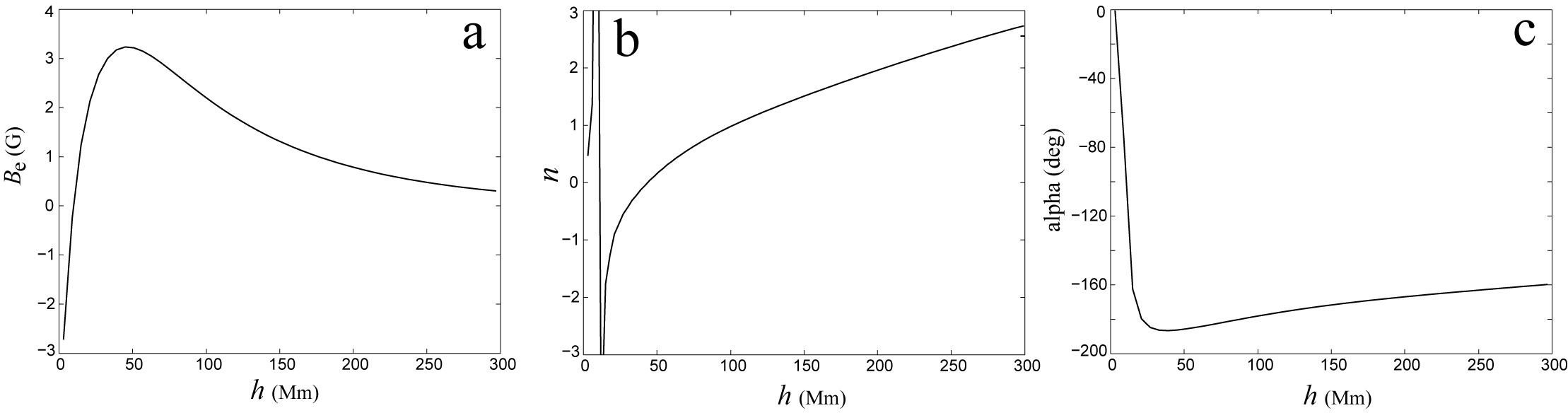}
    \caption{Dependence on height of the horizontal potential magnetic field $B_e$ (a), the decay index $n$ (b), the rotation angle $\alpha$ of $B_e$ (c) above the middle of the previous filament position on 2013 August 7. }
    \label{Fig6}
\end{figure*}

\begin{figure*}
		\includegraphics[width=180mm]{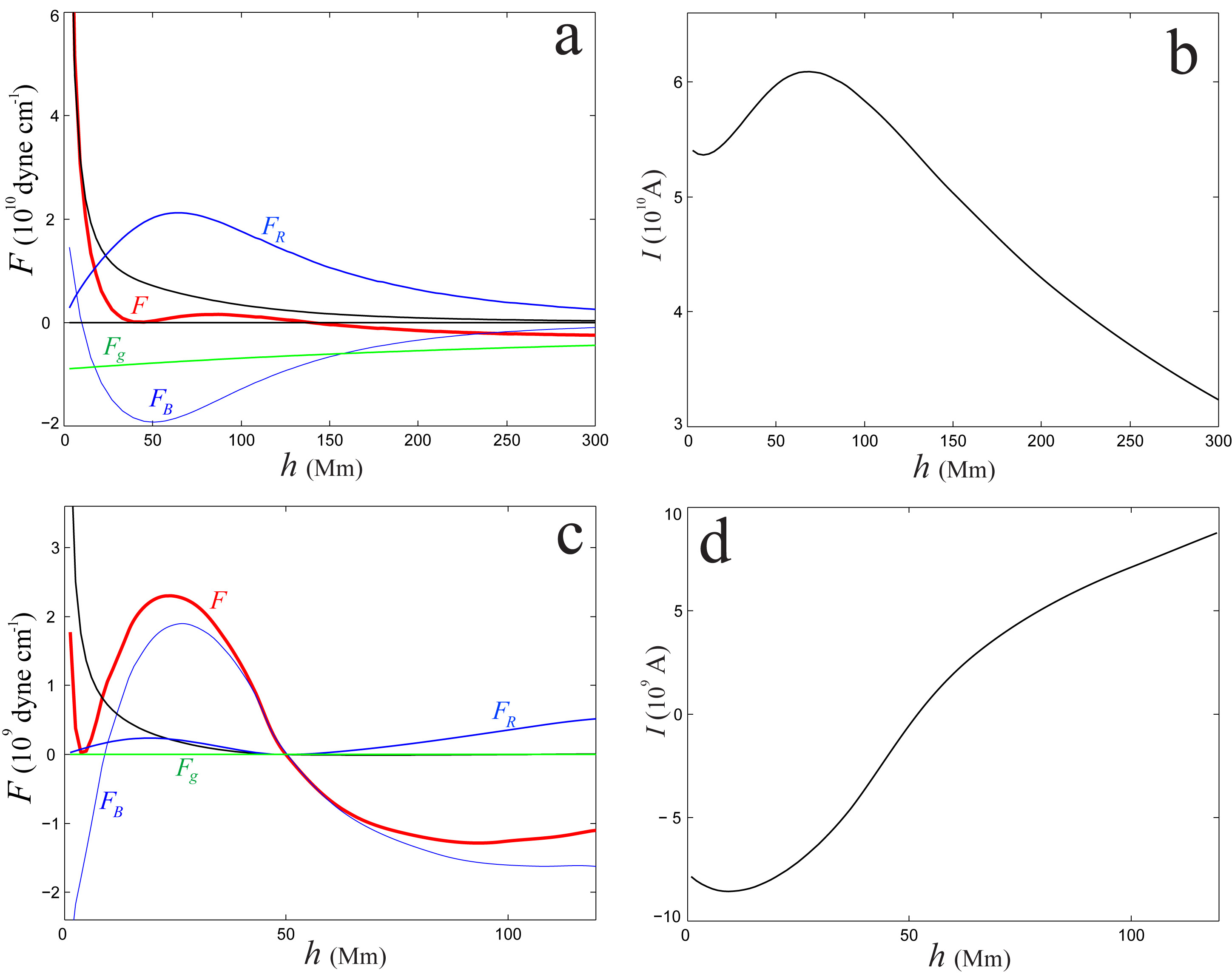}
    \caption{Dependence on height of the forces acting on the flux rope on 2013 August 7 (a, c) and the value of electric current $I$ in the flux rope (b, d) for opposite directions of the electric current in the unstable equilibrium point. }
    \label{Fig7}
\end{figure*}

\section{Results}
The described procedure was applied to the analysis of 15 failed eruptions. Since some eruptions happened close to the limb, magnetograms on a day, when a source region was close to the central meridian, were used for calculations. According to characteristic properties of the coronal magnetic field, the results are presented in two tables. In Table 1, events that happened in the regions with the monotonic decay-index height-dependence of the ambient magnetic field are listed. Table 2 presents events in the regions with the non-monotonic decay-index height-dependence. Values of the horizontal potential magnetic field, perpendicular to the plane of the flux rope, at the height of about 10 Mm halfway between the flux-rope endpoints are demonstrated in the eighth column. The ninth column shows the values of the rotation angle of the potential magnetic field $\alpha$ at the height $h_m$ comparative to the initial direction near the photosphere. The total mass of a filament $M$ is displayed in the last column.

The mass needed to terminate the eruption in the monotonic decay-index magnetic field could be expected to be less then in the non-monotonic decay-index field, because the field acts downwards on the flux rope at any height, while in the latter case it can act upwards due to the change of direction to opposite. However, the values of mass in Table 1 are systematically greater than in Table 2. One reason can be the strength of the external field $B_e$, which is typically greater in Table 1. The correlation between the field strength, electric current, and filament mass is evident. In most cases, the value of the filament mass (of the order of 10$^6$ g cm$^{-1}$) seems reasonable, corresponding to the ion number density of the order of $10^{11}-10^{12}$ cm$^{-3}$ in filaments. Nevertheless, in several events (on 2013.05.21, 2014.03.28, 2014.02.17) the mass is too large. 

For comparison, we considered six successful eruptions in regions with two different types of the decay-index behaviour (Tables 3 and 4). We can find only the upper limit of the filament mass assuming the upper equilibrium at the height of about the solar radius $R_\odot$. The values of the upper estimates of the mass are similar to the values obtained for the failed eruptions. The magnitudes of the magnetic field and critical electric current are also similar to those of the failed eruptions. Masses of filaments needed to terminate these eruptions should be much greater than the upper estimates shown in Tables 3 and 4. 

One event falls out of line. This is the failed eruption on 2013 August 04. The magnetic field, although it can be analysed only on August 07 when the region comes closer to the central meridian, changes direction showing the presence of a null point at a very low height of about 10 Mm (Figure 6), in contrast to the other analysed regions. Formally, we can derive the reasonable mass of the filament shown in Table 2, but the point of the loss of equilibrium is too high (about 40 Mm, above the null point) unlike the clearly observed low initial height of the filament of about 10 Mm [Figure 7(a)]. However, there is another possibility for unstable equilibrium below the null point with the opposite initial direction of the electric current [Figure 7(c)]. In this case, the current changes direction to opposite during the eruption due to inductance [Figure 7(d)]. Strong current changes lead to termination of the eruption at a rather low height of about 50 Mm even in the absence of gravity. Unfortunately, this height is significantly lower than the observed maximum height of the filament. Therefore, our model cannot fit the observed parameters of the eruption on 2013 August 04. Possibly, additional agents acting on the filament should be taken into account \citep{My15,Gr16,Fi20a}.

\section{Discussion  and conclusions}

 We estimated the mass of eruptive prominences using the model of a partial current-carrying torus loop anchored to the photosphere. The gravity force seems the most suitable agent to stop the eruption at a height comparable with the initial length of the eruptive filament. The obtained estimates of mass range from $4\times10^{15}$ to $270\times10^{16}$ g. For most events, the masses are in accordance with the measurements based on spectroscopic and white-light methods. In several events, the mass is much larger that it is usually expected in prominences. However, in the failed eruptions we deal with rather short and low laying prominences, which are not too convenient targets for a detailed spectroscopic study. High quiescent prominences are more suitable for this purpose and since plasma within them is more rarefied than in lower lying active-region and intermediate filaments, the results in optics could show lower magnitudes than obtained from the balance of forces. 
 
 The estimations of the mass of three filaments (namely, observed on 2013 May 21, 2014 March 28, 2014 February 17) much over $10\times10^{16}$ g seem not too reliable. It looks that some other forces not taken into account in our model play significant roles in the termination of eruptions of these filaments. \citet{Gr16}  pointed out several various mechanisms that can decelerate and stop eruptions in the corona. The first one is the action of gravitation on a filament, which possesses the amount of energy not enough to escape the gravitational potential of the Sun. This is the mechanism considered in this paper. 
 
 The second cause is the pressure of overlying magnetic field \citep{To05,Wa07,Am18}. The amount of stored free magnetic energy in the region before the eruption can be also not sufficient to overcome coronal magnetic confinement \citep{Liu18}. We consider only the potential field in the corona. While it also can be rather peculiar as in 2013 August 04 event, in all three ‘overloaded’ events the potential field structure looks not too different from others. However, the field strength and consequently the initial value of the flux-rope electric current are significantly greater. We can also underestimate the value of the magnetic field at greater heights because we use only limited area of the photosphere for the boundary condition in solving the Neumann external boundary-value problem, while distant magnetic concentrations may add to the field strength at great heights. There is also possibility that the coronal magnetic field, or a ‘magnetic cage’ for an eruptive flux rope, may be strengthen by field-aligned electric currents, which add the excess of a force-free field over the potential field \citep{Am18}. 

It is widely accepted that the strapping magnetic field plays the crucial role in the prospects of a filament eruption to be successful or failed. Sometimes the strapping magnetic field is manifested by a EUV arcade \citep{Ch13}. Many authors analyse whether the decay index in the interesting height range is below the threshold for the torus instability ($n_c = 1.5$) or above it. Failed filament eruptions on 2005 May 27 \citep{Gu11}, 2011 March 9 \citep{Li18}, and 2014 April 7 \citep{Xu16} were observed at heights where the decay index was below the threshold. However, sometimes eruptive filaments stop at heights where the decay index is greater that the threshold for the torus instability. It happens in most of our events in the regions with the monotonic decay index behaviour. In the regions with the non-monotonic decay-index height-dependence, the distribution of the index values at the final height is rather wide, including values below the threshold, but the direction of the horizontal field at this height is turned through more than 100$^\circ$ relative its direction at the initial height. This rotation of the coronal-field direction was not taken into account in many studies. On the other hand, \citet{Zh19} relate the failure of filament eruptions with strong writhing motions during the eruptions with the rotation of the filament axis in the range of 50$^\circ$ — 130$^\circ$.

Magnetic tension within the erupting flux rope is considered as a restricting force for its motion \citep{Vr90}. \citet{Ji03} argued that observed in the failed filament eruption on 2002 May 27 deceleration, exceeding the gravitational deceleration by a factor of ten, suggests that the filament material is pulling back by magnetic tension. We take into account the magnetic tension in our model, but for force-free internal structure of the flux rope the tension cannot overcame the hoop force stretching the curved flux-rope axis, but only reduces it. However, if the internal field is not force-free, the tension may play more significant role. For example, enlarged coronal plasma pressure can keep a greater longitudinal field than in the force-free case resulting in the increased tension. 

Laboratory experiments \citep{My15}  showed that the presence of a strong toroidal magnetic field along the flux-rope axis in the ambient space can prevent a flux rope eruption. This guide field interacts with the flux-rope electric current and produces a dynamic toroidal field tension force that is able to terminate the eruption. Precise measurements of forces acting on low-aspect-ratio, line-tied magnetic flux ropes \citep{My16}  revealed that the hoop force is systematically smaller than predicted by analytical expressions for large-aspect ratio flux ropes. On the other hand, the toroidal field tension force in the laboratory experiments is larger than it is expected from the analytical theory. Both this factors reduces the role of gravity in preventing successful eruptions.

Kink instability is widely believed to be able to destabilize a flux rope but it does not lead itself to a full successful eruption if the magnetic field above the flux rope does not decrease with height sufficiently steep\citep{To05,Fa07,Fa10}. In our events, the overlaying potential field changes with height rather fast and in many cases it changes its direction to nearly opposite to the initial direction. Moreover, there is no significant rotation of the filament arch plane observed during the ascent in all studied failed eruptions. Therefore, there is no evidence of writhing and kink-instability in this sample of events, although we cannot rule out absolutely the kink-instability as a trigger of eruptions.

\citet{Am99} presented the results of the three-dimensional numerical simulations where a twisted flux rope is destabilized and after a phase of dynamic evolution disrupts into two almost untwisted flux tubes confined within a closed overlaying arcade. During the relaxation of the magnetic configuration, the magnetic field of the flux tube reconnects with the overlaying arcade, and some part of the free magnetic energy is released. The model shows some possible scenario of a failed flux-rope eruption and confined flare. Reconnection of coronal field lines above filaments before eruptions can either strengthening or weakening the magnetic confinement of filaments facilitating either failed or successful eruptions\citep{Wa18}). 

The existence of a high-altitude stable equilibrium point is only a necessary condition for a failed eruption. There should be a drag force that prevents the flux rope from large-amplitude and long-time oscillation about the high equilibrium position. In the absence of the drag force, the flux rope could reach too high height where the strapping external field becomes too small to bring the flux rope back to the equilibrium point. In other words, if the kinetic energy of the eruptive filament does not dissipate during the ascending motion, it could be enough to support a successful eruption. 

In the illustrative example shown in Figure 5, we used two functional dependences of the drag force on velocity. The linear dependence is usually accepted in hydrodynamics for small values of the Reynolds number 
\begin{equation}
\mathrm R =  \frac{\rho v l}{\eta} ,
\end{equation}  
 where $\rho$ is the density, $l$ is a characteristic size, and $\eta$ is the coefficient of dynamic viscosity. For great values of the Reynolds number, the drag force is proportional second degree of velocity. We chose the values of the coefficients of proportionality rather arbitrarily as $k_1 = 1.8\times10^4$ g cm$^{-1}$ s$^{-1}$, $k_2 = 1.1\times10^{-2}$ g cm$^{-2}$ in order only to damp oscillations of the flux rope about the upper equilibrium position. Evidently, at the beginning of eruption the velocity is low, and the Reynolds number is small. After the main phase of acceleration, the velocity is of the order of $10^7$ cm s$^{-1}$. Assuming the typical coronal density $\rho = 2\times10^{-15}$ g cm$^{-3}$, $l = 2r_0 = 10^9$ cm, and $\eta \approx 1$ g s$^{-1}$ cm$^{-1}$, \citep{Za18}  we obtain R = 20. \citet{Za18} considered motion of a smaller flux rope with a lower speed. The Reynolds number for their case is about unity. They found $k_1 \approx 10$ g cm$^{-1}$ s$^{-1}$, which is three orders less than used in our calculations. The corresponding decay time of the flux-rope oscillation was in their estimations about 10 hours. Of course, it is too long for failed eruptions. True enough, they use the expression for the drag force action on a long cylinder valid strictly only for R $\ll$ 1. In eruption of filaments with lesser mass, the coefficient $k_1$ is also needed lesser, but still greater than typical for low-Reynolds-number motion.

Expression for the drag force for large Reynolds-numbers has the form (Landau and Lifshitz 1987)
\begin{equation}
F_d = - C \rho S v^2 ,
\end{equation}  
where $C$ is the dimensionless coefficient dependent on the Reynolds number and the body shape and $S$ is the cross-section area of the body. The value of $C$ decreases from $\approx$ 100 to $\approx$ 0.5, when R increases from 0.1 to 1000. For the eruption described in Section 4, $k_2 = C \rho S \approx  2\times10^{-5}$ g cm$^{-2}$. This value is also three orders less than used in our calculations. 

Wide discrepancy of the drag coefficients needed to damp oscillations of the flux rope about the upper equilibrium position with the theoretical estimations can be for reasons of inapplicability of hydrodynamic formulas derived for solid bodies in viscous liquid for coronal conditions or presence of additional dissipative processes. Anyway, our kinematics calculations have only illustrative character. 

Although the filament mass seems to be the major factor able to terminate eruptions of rather short filaments, the example of the eruption on 2013 August 04 shows that there are other possibilities. The specific distribution of the coronal magnetic field can induce strong changes of the flux-rope electric current due to inductance, which can turn over the action of the external field from restraining to pushing upwards and backwards [Figure 7(c, d)]. There can be also other mechanisms of confining eruptions.

\begin{acknowledgements}
The author thanks the Big Bear Solar Observatory, the Kanzelhohe Solar Observatory, the Udaipur Solar Observatory, the {\it SOHO}, the {\it STEREO}, and the {\it SDO} scientific teams for the high-quality data they supply. {\it SOHO} is a project of international cooperation between ESA and NASA. STEREO is the third mission in NASAs Solar Terrestrial Probes program. {\it SDO} is a mission of NASAs Living With a Star Program. This work utilizes GONG data from NSO, which is operated by AURA under a cooperative agreement with NSF and with additional financial support from NOAA, NASA, and USAF.

\end{acknowledgements}

\bibliographystyle{pasa-mnras}
\bibliography{Reference}

\end{document}